\documentclass[%
 reprint,
 amsmath,amssymb,
 aps,
pra,
]{revtex4-1}

\usepackage{graphicx}
\usepackage{dcolumn}
\usepackage{bm}

\begin{document}
\preprint{APS/123-QED}
\title{Measuring the hyperfine splittings and deriving the hyperfine-interaction \\constants of Cesium 7D$_{5/2}$ excited state}
\author{Zerong Wang,${}^{1}$ Xiaokai Hou,${}^{1}$ Jiandong Bai,${}^{1}$}
\author{Junmin Wang${}^{1,2,3,}$}
\email{wwjjmm@sxu.edu.cn}
\affiliation{{${}^{1}$State Key Laboratory of Quantum Optics and Quantum Optics Devices, and Institute of Opto-Electronics, Shanxi University, Tai Yuan 030006, Shanxi Province, People’s Republic of China}\\{${}^{2}$Department of Physics, School of Physics and Electronic Engineering, Shanxi University, Tai Yuan 030006, Shanxi Province, People’s Republic of China}\\{${}^{3}$Collaborative Innovation Center of Extreme Optics of the Education Ministry and Shanxi Province, Shanxi University, Tai Yuan 030006, Shanxi Province, People’s Republic of China}}
\date{\today}
\begin{abstract}The measurement of  Cesium (Cs) 7D$_{5/2}$ excited state's hyperfine splitting intervals and hyperfine-interaction constants has been experimentally investigated based on ladder-type (852 nm + 698 nm) three-level Cs system (6S$_{1/2}$ - 6P$_{3/2}$ - 7D$_{5/2}$)with room-temperature Cs atomic vapor cell. By scanning the 698-nm coupling laser’s frequency, the Doppler-free high-resolution electromagnetically-induced transparency (EIT) assisted double-resonance optical pumping (DROP) spectra have been demonstrated via transimission enhancement of the locked 852-nm probe laser. The EIT-assisted DROP spectra are employed to study the hyperfine splitting intervals for the Cs 7D$_{5/2}$ excited state with a room-temperature cesium atomic vapor cell, and the radio-frequency modulation sideband of a waveguide-type electro-optic phase modulator(EOPM) is introduced for frequency calibration to improve the accuracy of frequency interval measurement. The existence of EIT makes the DROP spectral linewidth much narrower, and it is very helpful to improve the spectroscopic resolution significantly. Benefiting from the higher signal-to-noise ratio (SNR) and much better resolution of the EIT-assisted DROP spectra, the hyperfine splitting intervals between the hyperfine folds of (\emph{F}" = 6), (\emph{F}" = 5), and (\emph{F}" = 4) of cesium 7D$_{5/2}$ state (HFS$_{ 6"-5"}$ = -10.60(0.17) MHz and HFS$_{5"-4"}$ = -8.54(0.15) MHz) have been measured, and therefore the magnetic-dipole hyperfine-interaction constant (\textbf{\emph{A}} = -1.70(0.03) MHz) and the electric-quadrupole hyperfine-interaction constant (\textbf{\emph{B}} = -0.77(0.58) MHz) have been derived for the Cs 7D$_{5/2}$ state. These constants have important reference value for the improvement of precise measurement and determination of basic physical constants.
\end{abstract}

\maketitle


\section{Introduction}
\textit{}
In recent years, with the maturity and development of high-precision spectroscopy technology, the precise measurement of the hyperfine structure (HFS) of alkali metal atoms and related physical constants has been one of concern topics in the fields of atomic physics, laser spectroscopy, and precision measurement [1, 2]. To reduce the Doppler broadening effect, two-photon Doppler-free spectroscopy [3] and EIT [4] has been used for precise measurements. The hyperfine splitting intervals of the excited state for alkali metal atoms can be measured and the corresponding hyperfine-interaction constants can be further derived with high precision. Hyperfine structure of atoms which resulted from the electron-nuclear interactions provides the structural information about the nuclear and electronic structure of the atoms [5]. There is a further problem with the parity non-conservation (PNC) [6, 7]. The PNC is sensitively dependent on the overlap of the nuclear and the electronic wave function [8] which like the hyperfine-interaction constants is related to accurate atomic structure. Owing to hyperfine structure can be accurately measured,  we can improve the accuracy and accuracy of measurements to explain the relevant physical the measurements and calculations [9, 10]. B. K. Sahoo and B. P. Das [11] recently studied calculation of the nuclear spin-independent parity violating electric dipole transition amplitude and constraints on new physics from an improved calculation of parity violation in Cs. On the other hand, another challenging problem which arises in this domain is that Safronova and Clark [12] found that the polarizability of 6P state is consistent with the experimental values, but the lifetime is inconsistent with the experimental values of 5D state. The polarizability and the lifetime are also closely related to the calculation of the dipole matrix element [10, 13]. In order to analyse the origin of this inconsistency, it is necessary to continue to study the structural properties of nD states. V. A. Dzuba et al [14] who studied the feasibility of measuring PNC amplitudes in the dipole-forbidden transitions of cesium are hampered by the difficulty of the strong correlation effects. In order to solve this problem, he is still calling for studying the nD hyperfine structure. In summary, the precise measurement of hyperfine structures plays an important role in fundamental physics such as atomic frequency calibration, construction of quantum theoretical models, laser cooling and trapping, and isotope identification [15-19].

Although many groups have carried out theoretical and experimental studies on the atomic hyperfine structure, there are few precise measurements of the hyperfine-interaction constants, especially for the nD state. Considering the complexity of its internal electronic cloud structure and strong correlation effects [14], the calculation results of different theoretical models differ greatly. In order to verify the theoretical model of nD state, it is urgent to accurately measure the hyperfine level structure of D state in the experiment. In 1995, T. T. Grove et al [20] measured the hyperfine interaction constant of $^{85}$Rb 5D$_{5/2}$ state by using optical double resonance spectroscopy and performed frequency calibration using an acousto-optic modulator (AOM). M. Auzinsh et al [21] conducted experimental and theoretical research on the polarizability and hyperfine-interaction constants of Cs atoms in D state, and the experimental value of the hyperfine interaction constants were obtained from the energy level crossover signal of D state. Based on the atomic coherence effect of 6S$_{1/2}$ - 6P$_{3/2}$ - 8S$_{1/2}$ system in cesium atom cell at room temperature, J. Wang et al [22] proposed a new high-resolution technique for measuring the hyperfine division of the excited state and developed a method to eliminate error arising from the nonlinear frequency scanning by employing an optical waveguide-type electro-optic phase modulator and a confocal Fabry–Perot cavity [23]. Using this technique, G. Yang et al [24] measured the hyperfine-interaction constants of 5D$_{5/2}$ state. Recently, Y. H. He et al [25] presented a precise measurement of the hyperfine structure of cesium 7S$_{1/2}$ excited state by using electro-magnetically induced spectroscopy with a cesium three-level cascade system. J. P. Yuan et al [4] studied the electro-magnetically induced grating diffraction controllability in a ladder-type (5S$_{1/2}$–5P$_{3/2}$–5D$_{5/2}$) rubidium atomic system and the nonlinear dependence of the first-order diffraction efficiency on the coupling laser power for the first time.

As for 7D$_{5/2}$ state, only a few studies have shown relevant measurement of its structure. B. R. Bulos et al \cite{ref-journal26} determined the \textbf{\emph{A}} value of cesium 7D$_{5/2}$ state by optical double resonance transition spectra. Y. C. Lee et al [27] used cesium 6S$_{1/2}$ - 7D$_{5/2}$ two-photon transition method by introducing an electro-optical phase modulator to determine the corresponding hyperfine interval and hyperfine-interaction constants of cesium 7D$_{5/2}$ state. J. E. Stalnaker et al [28] measured the absolute transition frequency and the hyperfine-interaction constants of cesium 7D$_{5/2}$ states by using the frequency comb method in the two-photon excitation system. Recently, Sandan Wang[29] used a single laser with 767 nm to study the 6S$_{1/2}$ - 7D$_{3/2, 5/2}$ electric quadrupole transition in a thermal Cs vapor. The magnetic dipole coupling constant \textbf{\emph{A}} and electric quadrupole coupling constant \textbf{\emph{B}} for the 7D$_{3/2, 5/2}$ states are precisely determined by using the hyperfine splitting intervals. 

Due to the fact that its adjacent splitting intervals are less than $\sim$10 MHz, the main practical problem that confronts us is how to improve the spectroscopic resolution and (signal-to-noise ratio) SNR of 7D$_{5/2}$ state. Compared with the previous experiments and theoretical studies, we used the ladder-type EIT which was employed to improve the spectroscopic resolution and SNR. We used 698-nm coupling laser and 852-nm probe laser in opposite to realize the Cs cascade two-photon excitation in Cs atomic vapor cell at room temperature. Also the effect of 852-nm probe beam’s linewidth and optical intensity of 852-nm probe beam and 698-nm coupling beam upon the spectroscopic SNR were investigated. After study the effect of spectroscopic scheme and optimization of experimental parameters, the spectrum with higher SNR and better resolution EIT-assisted DROP signals were observed, and the hyperfine splitting intervals of Cs 7D$_{5/2}$ state were measured. The sideband of an optical waveguide-type EOPM was introduced as a frequency calibration to improve the accuracy of frequency measurement. The hyperfine splitting intervals between the hyperfine folds of (\emph{F}" = 6), (\emph{F}" = 5), and (\emph{F}" = 4) of Cs 7D$_{5/2}$ state (HFS$_{ 6"-5"}$ = -10.60(0.17) MHz and HFS$_{5"-4"}$ = -8.54(0.15) MHz) have been measured, and therefore the magnetic-dipole hyperfine-interaction constant (\textbf{\emph{A}} = -1.70(0.03) MHz) and the electric-quadrupole hyperfine-interaction constant (\textbf{\emph{B}} = -0.77(0.58) MHz) for the cesium 7D$_{5/2}$ state have been derived.

The paper is organized in five sections. In Sec.II, we introduce relevant principles including the hyperfine splitting and the momentum quantum number, the hyperfine splitting and the hyperfine-interaction constants, and the spectroscopic schemes we choose in this work. In Sec.III we describe the experimental setup and conditons under which we performed the measurements. In Sec.IV presents optimization of experimental parameters in order to get the higher SNR and much better resolution of the EIT-assisted DROP spectra. Then we measured the hyperfine splitting intervals and the hyperfine interaction constants of 7D$_{5/2}$ state more precisely by analyzing the experimental data and discussing the related techniques in Sec.V. and then outline the main points in Sec.VI.

\section{Relevant principles}
\subsection{The hyperfine splitting and the momentum quantum number}
Hyperfine interaction includes magnetic-dipole hyperfine interaction and electric-quadrupole hyperfine interaction. In general, the hyperfine splitting is mainly considered from the following two aspects: nuclear spin of atoms and total angular momentum of electrons, leading to atomic energy level movement [30]; The interaction between the electric field generated by the external valence electrons and the electric-quadrupole moment also causes the energy level to shift. Therefore, the total shift of energy level can be expressed as follows [24]:

\begin{eqnarray}
{{\Delta}{E}}=\frac{1}{2}{\textbf{\emph{A}}}{K}+{\textbf{\emph{B}}}\frac{\frac{3}{2}{K(K+1)-2I(I+1)J(J+1)}}{4I(2I-1)J(2J-1)}, \quad 
\end{eqnarray}
where \emph{K}=\emph{F}(\emph{F}+1)-\emph{I}(\emph{I}+1)-\emph{J}(\emph{J}+1) is the total angular momentum quantum number of the nucleus, \emph{J} is the total angular momentum quantum number of the electron, \emph{F}=\emph{I}+\emph{J} is the total angular momentum quantum number of atoms, \textbf{\emph{A}} and \textbf{\emph{B}} are magnetic-dipole hyperfine-interaction constant and electric-quadrupole hyperfine-interaction constant, respectively.
\vspace{0.05in}

\subsection{The hyperfine splitting and the hyperfine-interaction constants}
From formula (1), it can be deduced that the adjacent hyperfine interval is:

\begin{equation}\label{eq:2}
\begin{split}
& {{\Delta}{E}(\emph{F}\rightarrow\emph{F}-1)}=\\
& {\textbf{\emph{A}}}{F}+{\textbf{\emph{B}}}\frac{\frac{3}{2}{\emph{F}\,[\emph{F}^{\,2}-\emph{I}(\emph{I}+1)-\emph{J}(\emph{J}+1)]}}{\emph{I}(2\emph{I}-1)\emph{J}(2\emph{J}-1)},\qquad
\end{split}
\end{equation}

For S state, the orbital angular momentum L = 0, the electronic wave function is spherically symmetric distribution. Due to the nucleus electric field gradient is zero, there is only the magnetic dipole hyperfine effects without electric quadrupole effects. The magnetic-dipole hyperfine constant can be calculated directly from  formula (2). But for D state, the calculation of the hyperfine splitting of \textbf{\emph{A}}, \textbf{\emph{B}} can be obtained from the equation, according to the formula (2) [31]. 

\subsection{The spectroscopic schemes}
High-resolution and high SNR spectroscopic schemes are very helpful to achieve higher accuracy of hyperfine splitting measurement, so as to deduce accurate hyperfine-interaction constants. (here, the magnetic-dipole hyperfine-interaction constant \textbf{\emph{A}} and the electric-quadrupole hyperfine-interaction constant \textbf{\emph{B}} for the cesium 7D$_{5/2}$ excited state).

Double-resonance optical-pumping (DROP) spectroscopic scheme based on the ladder-type three-level system (6S$_{1/2}$ - 6P$_{3/2}$ - 7D$_{5/2}$) is employed to measure the hyperfine splitting in cesium 7D$_{5/2}$ excited state. 

\begin{figure}[b]
\vspace{-0.00in}
\centerline{
\includegraphics[width = 86mm]{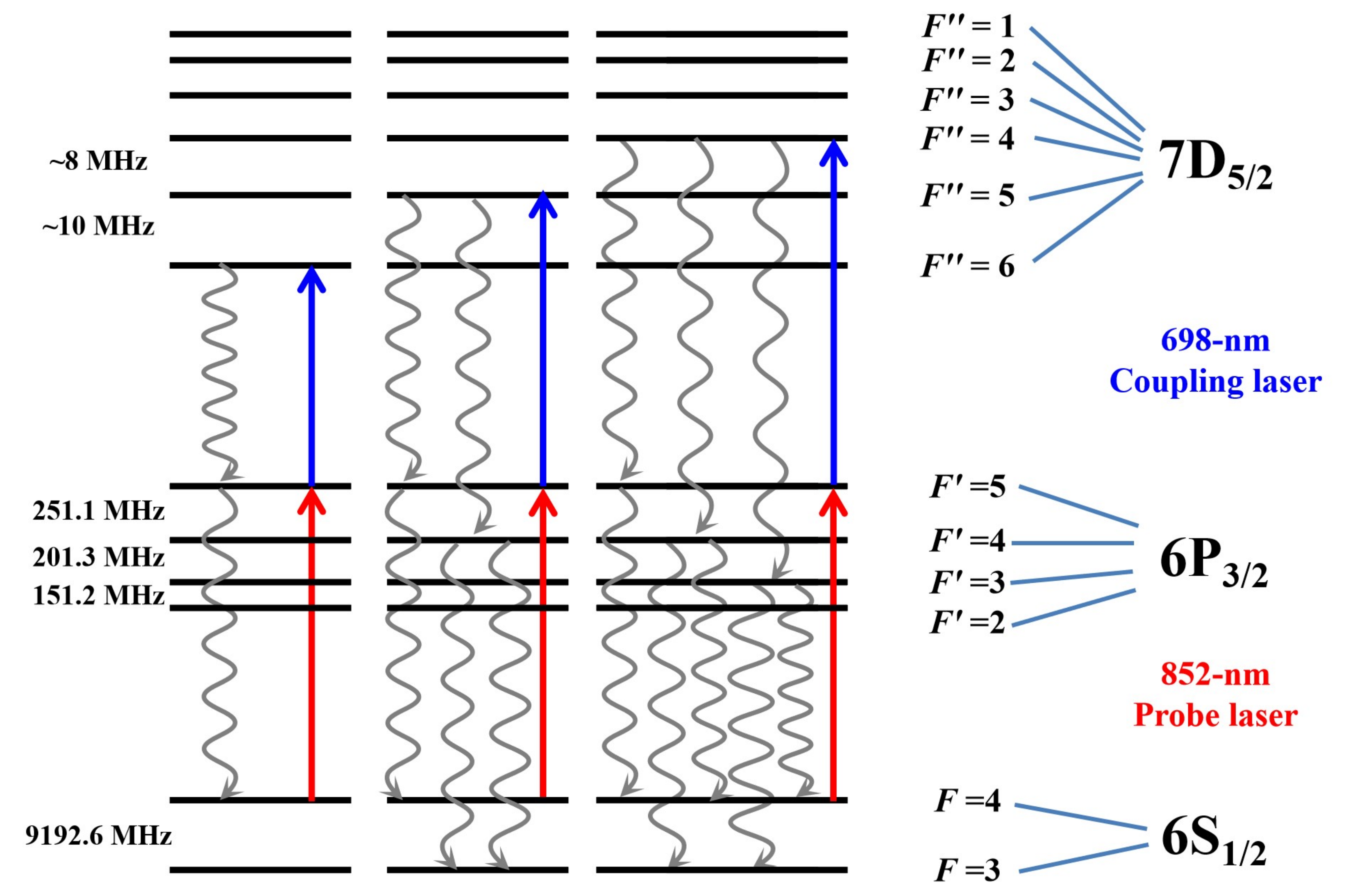}
} \vspace{-0.1in} 
\caption{The relevant hyperfine energy levels of cesium atoms. Double-resonance optical-pumping (DROP) spectroscopic scheme based on the ladder-type three-level system (6S$_{1/2}$ - 6P$_{3/2}$ - 7D$_{5/2}$) is employed to measure the hyperfine splittings of Cs 7D$_{5/2}$ excited state. The 852-nm probe laser drives the 6S$_{1/2}$ (\emph{F} = 4) - 6P$_{3/2}$ (\emph{F}' = 5) transition which is indicated by the red arrows, while the 698-nm coupling laser drives the 6P$_{3/2}$ (\emph{F}' = 5) - 7D$_{5/2}$ (\emph{F}" = 6), (\emph{F}" = 5) and (\emph{F}" = 4) transitions which are indicated by the three blue arrows respectively. The wavy arrows indicate the spontaneous decay channels.}
\label{Fig 1}
\vspace{-0.0in}
\end{figure}

An 852-nm laser is used as the probe laser to detect the changes of the population of ground state, which reflecting the hyperfine spectrum between excited states. At the same time, we also compare the Optical-optical double-resonance (OODR) absorption spectroscopic scheme and use 698-nm laser as the probe laser to detect the changes of the population of intermediate atoms to reflect the hyperfine spectrum between excited states. OODR spectroscopic signal is related to the population between 6P$_{3/2}$ (\emph{F}' = 5) and 7D$_{5/2}$ (\emph{F}" = 4, 5, 6), while 6P$_{3/2}$ (\emph{F}' = 5) has a large spontaneous decay rate ($\Gamma$ = 2$\pi$$\times$5.22 MHz), so atoms are not easy to inhabit in 6P$_{3/2}$ (\emph{F}' = 5) state. On the contrary, it will accelerate the DROP process, which will rapidly reduce the population of atoms on the ground state \emph{F} = 4, so its SNR is relatively higher, which is helpful to measure the hyperfine splitting interval more accurately.

In the three-level system shown in Fig. 1, when the coupling laser and the probe laser are in resonance, under the weak laser approximation, the polarizability of the medium can be expressed as [32]: 

\begin{widetext}
\begin{equation}
{\chi(\upsilon)\mathrm{d}\upsilon} = {\frac{4i\hbar g_{21}^2/\varepsilon_0}{\gamma_{21}-i\Delta_p-i\frac{\omega_p}{c}\upsilon+\frac{\Omega_c^2/4}{\gamma_{31}-i(\Delta_p+\Delta_c)-i(\omega_p\mp \omega_c)\upsilon/c}}}{N(\upsilon)\mathrm{d}\upsilon}
 \label{eq:wideeq},\quad
\end{equation}
\end{widetext}
where $\omega_{\emph{p}}$ is the frequency of probe laser, $\omega_{\emph{c}}$ is the frequency of coupling laser, $\Delta_{\emph{p}}$ is the detuning of probe laser relative to the frequency of the 6S$_{1/2}$ (\emph{F} = 4) - 6P$_{3/2}$ (\emph{F}' = 5) transition, $\Delta_{\emph{c}}$ is the detuning of coupling laser relative to the frequency of the 6P$_{3/2}$ (\emph{F}' = 5) - 7D$_{5/2}$ (\emph{F}" = 6) transition, the decay rate is ${\gamma_\emph{ij}} = ( \Gamma_{i}+\Gamma_{j})/2$, \emph{g}$_{21}$ is the dipole moment matrix element for the 6S$_{1/2}$ (\emph{F} = 4) - 6P$_{3/2}$ (\emph{F}' = 5) transition,  $\Omega_{\emph{c}} = 2{\gamma_{32}}{E_{\emph{c}}}$ is the Rabi frequency of coupling laser, \emph{$\Gamma$}$_{i}$ is the natural linewidth of level \emph{i}, \emph{N} is the density of Cs atoms, $\upsilon$ is the velocity of Cs atoms, $\emph{E}_\emph{c}$ represents the amplitude of the coupling laser field, \emph{c} is the speed of laser. In Equation(3), the term -\emph{i}($\omega_{\emph{p}}$ - $\omega_{\emph{c}}$)$\upsilon$/\emph{c} is corresponding to the counter-propagating (CTP) configuration, it is easy to achieve EIT signals in two-photon Doppler-free configuration, and the linewidth of the signal is narrow due to atomic coherence effect. For the co-propagating (CP) configuration, the term -\emph{i}($\omega_{\emph{p}}$ + $\omega_{\emph{c}}$)$\upsilon$/\emph{c} cannot be ignored. The Doppler background basically overwhelms the EIT signals [33].

In order to obtain the spectral signal with high resolution, we adopt the scheme that the probe and coupling beams counter-propagate through the Cs atomic vapor and scanning coupling laser’s frequency to obtain signals without Doppler background in the experiment.

\section{Experimental setup}
The experimental setup is shown in Fig.2. The 852-nm probe beam is provided by a tunable diode laser (external-cavity diode laser (ECDL) or distributed-Bragg-reflector (DBR) diode laser). Part of the laser output is used to lock the laser frequency to the cesium 6S$_{1/2}$ (\emph{F} = 4) - 6P$_{3/2}$ (\emph{F}' = 5) hyperfine transition by using of polarization spectroscopy (PS) scheme.
\begin{figure}[b]
\vspace{-0.00in}
\centerline{
\includegraphics[width = 86mm]{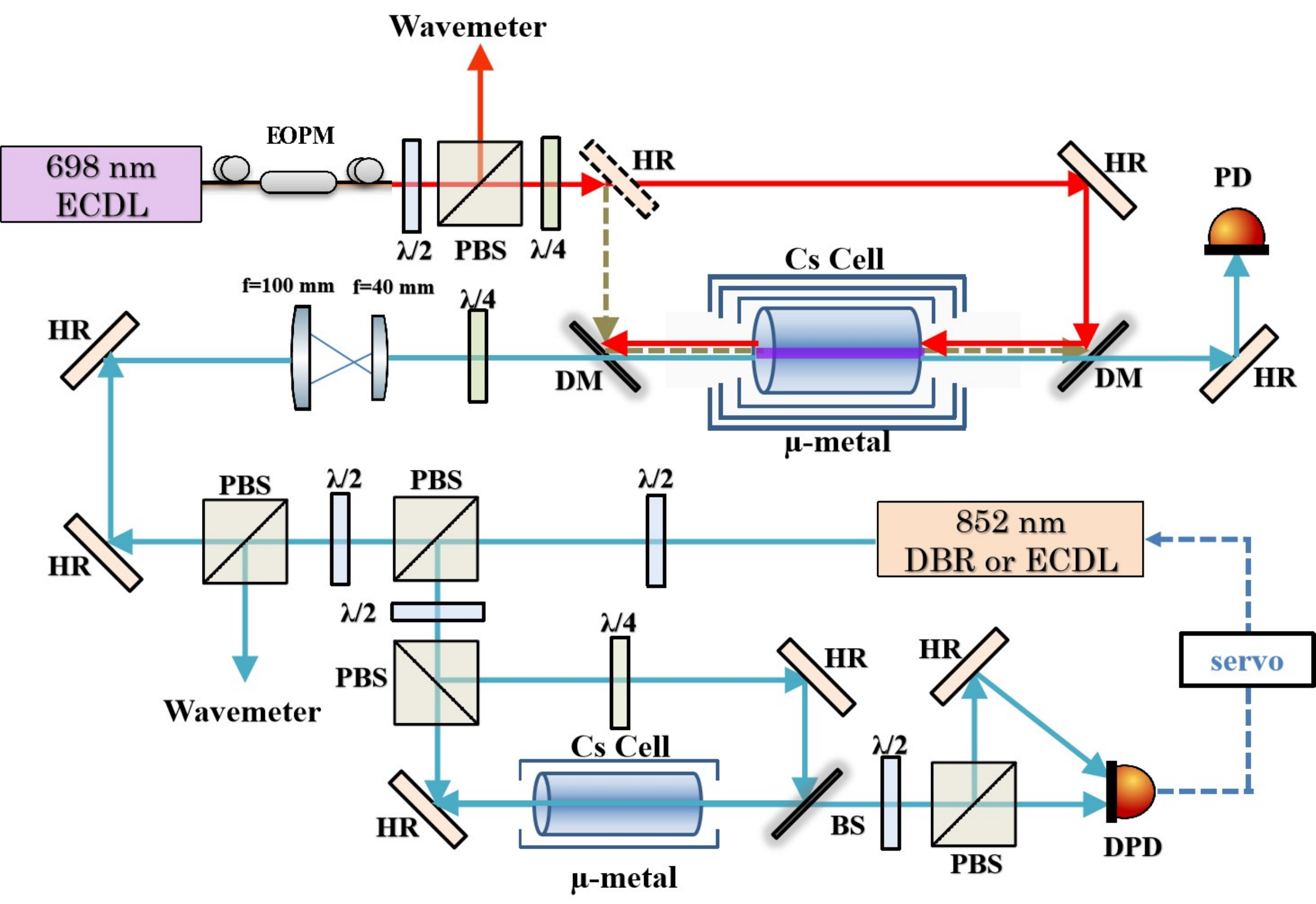}
} \vspace{-0.1in} 
\caption{Experimental setup for hyperfine splittings measurement: (red solid lines with arraws) counter-propagating (CTP) configuration with the probe and coupling laser beams and (green dash lines with arraws) co-propagating (CP) configuration. Where $\mu$-metal, magnetic shielding device; EOPM, waveguide-type electro-optic phase modulator; PBS, polarization beam splitter cube; BS: beam splitter plate; $\lambda$/2, half-wave plate; $\lambda$/4, quarter-wave plate; PD, photodetector; DPD: differential photodiode; DM, dichroic mirror; HR, high reflection mirror; servo, the frequency locking servo system.}
\label{Fig 2}
\vspace{-0.0in}
\end{figure}
The main 852-nm output laser beam overlaps with the 698-nm laser provided by an ECDL  with a typical linewidth of $\sim$200 kHz is transmitted through the cesium atomic vapor cell in the opposite direction of the 852-nm probe beam (CTP configuration). The 852-nm probe beam has 1/e$^{2}$ diameters of  $\sim$600 $\mu$m and the 698-nm coupling beam has 1/e$^{2}$ diameters of  $\sim$840 $\mu$m. The cesium vapor cell is wrapped with $\mu$-metal sheet in order to decrease the influence of background magnetic field. The wavelength meter (Advantest, TQ-8325) is employed to monitor the wavelength of 698-nm laser in real time.\\

\section{Optimization of experimental parameters}
\subsection{Comparison of OODR and DROP spectroscopic schemes}
In the experiment, we compare two spectroscopic signals by detecting different laser corresponding to OODR and DROP spectra respectively. The 852-nm probe laser transmits the atoms on ground state 6S$_{1/2}$ (\emph{F} = 4) into intermediate state 6P$_{3/2}$ (F' = 5). 698-nm coupling laser excites atoms into the 7D$_{5/2}$ (\emph{F}" = 4, 5, 6) state, scanned across the 6P$_{3/2}$ (F' =5) – 7D$_{5/2}$ transition. But some atoms on the 7D$_{5/2}$ (\emph{F}" = 4, 5, 6) state can be pumped to another ground 6S$_{1/2}$ (\emph{F} = 3) state through other intermediate states shown in Fig. 1, which leads to a change of the population due to the DROP. The resolution of DROP spectrum is better than OODR spectrum and there is basically no coincidence between the three peaks. It can be clearly seen from Fig. 3 that the resolution of DROP spectrum is better than OODR spectrum, which is helpful to measure the hyperfine splitting interval more accurately.
\begin{figure}[h]
\vspace{0.2in}
\centerline{
\includegraphics[width = 86mm]{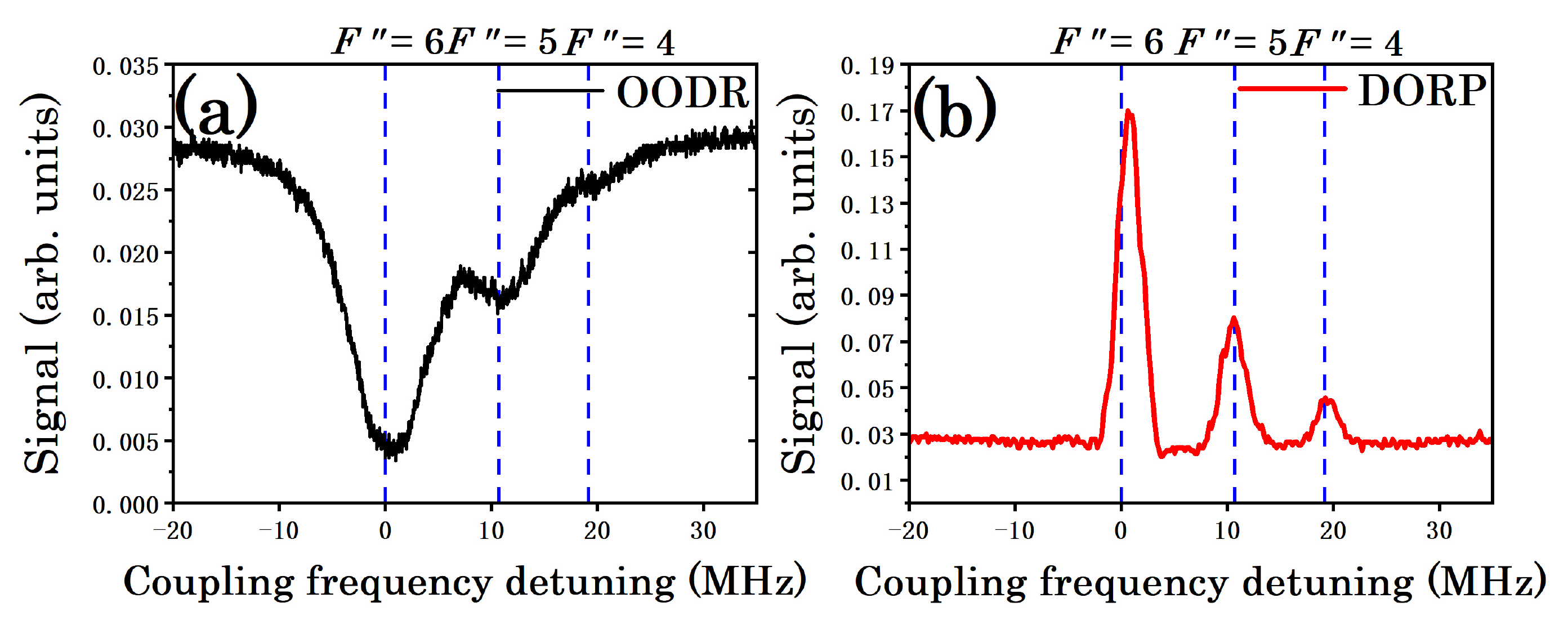}
} \vspace{-0.1in} 
\caption {Comparison of OODR and DROP spectroscopic signals for the cesium 6P$_{3/2}$ (\emph{F}' = 5)  - 7D$_{5/2}$ (\emph{F}" = 4, 5, 6) transitions. (a) OODR spectroscopic signal (b) DROP spectroscopic signal.}
\label{Fig 3}
\vspace{-0.0in}
\end{figure}

\subsection{Effect of the probe laser’s linewidth}

A possible solution to the problem at hand is that different linewidth of 852-nm lasers are used in the experiment, and the measured data of full-text are based on the ECDL with a typical linewidth of $\sim$200 kHz (except Fig.4 (a)). Similarly, we also compare a DBR diode laser with a typical linewidth of $\sim$2 MHz, and the spectra obtained is shown in Fig.4 (a). The difficulty in separating the three peaks suggests that the laser with a narrower linewidth will be very helpful to achieve a higher spectroscopic resolution. 
\begin{figure}[h]
\vspace{-0.00in}
\centerline{
\includegraphics[width = 86mm]{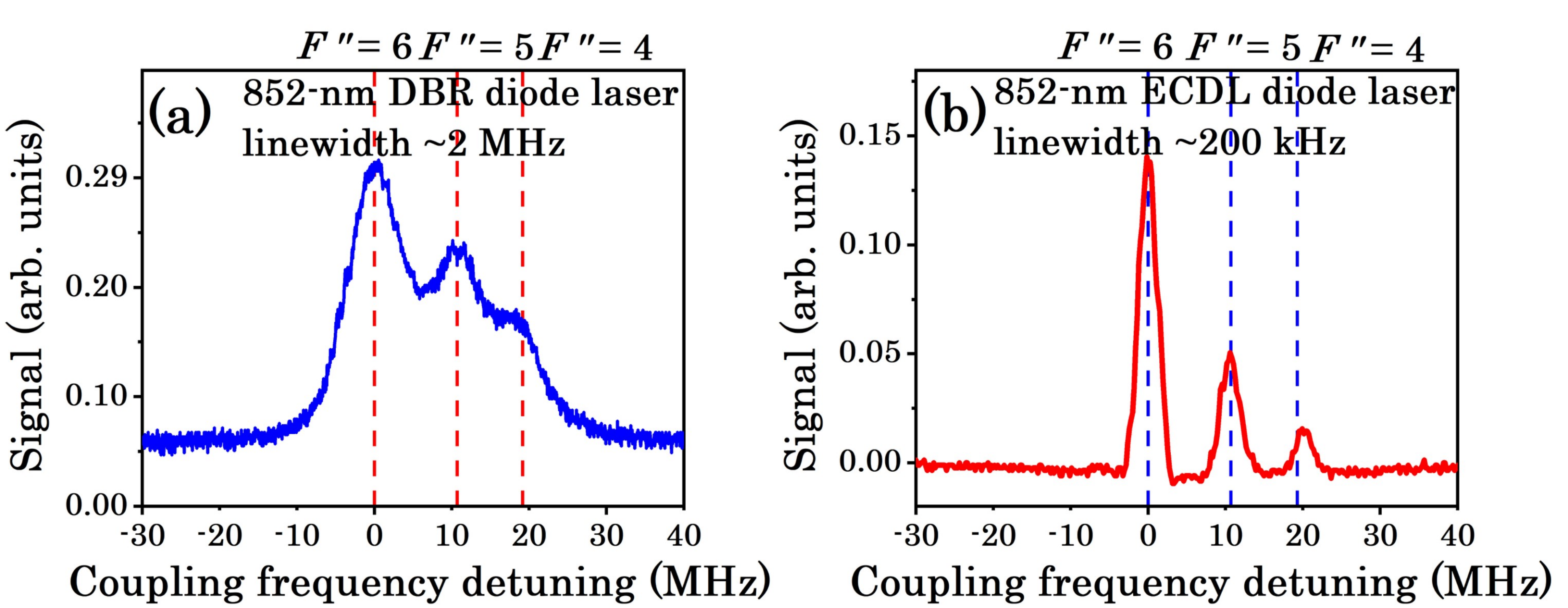}
} \vspace{0.1in} 
\caption {Two-photon transition spectra with no Doppler background are measured with two different 852-nm probe lasers. (a) An 852-nm Distributed-Bragg-reflector (DBR) type diode laser with a typical linewidth of $\sim$2 MHz; (b) An 852-nm ECDL with a typical linewidth of $\sim$200 kHz.}
\label{Fig 4}
\vspace{-0.0in}
\end{figure}

\subsection{EIT-assisted DROP spectra}
We compare the two transmission methods in the CP configuration and the CTP configuration; Two different scanning methods are used to obtain the spectral signals, as shown in Fig. 5, keeping the 698-nm coupling laser in free running, we obtain the two-photon transition spectra with Doppler background shown in Fig.5 (a) and (c) by scanning the 852-nm probe laser.

\begin{figure}[h]
\vspace{-0.00in}
\centerline{
\includegraphics[width = 86mm]{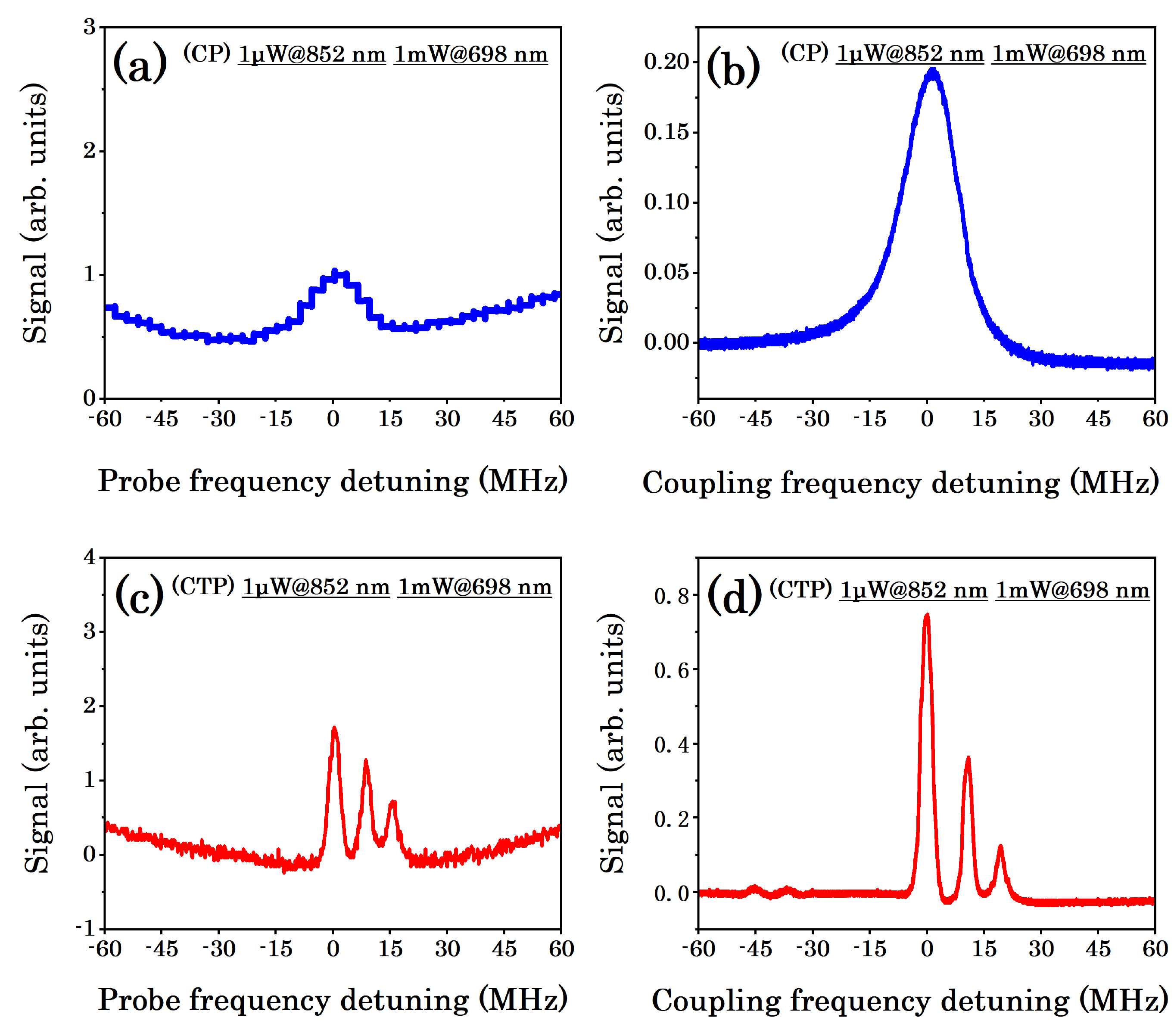}
} \vspace{-0.1in} 
\caption {Spectral signals under different transmission modes and scanning modes (a) and (b) in the CP configuration, and (c) and (d) in the CTP configuration. (a) and (c) The 698 nm laser was adjusted to the 6P$_{3/2}$(\emph{F}' = 5) - 7D$_{5/2}$(\emph{F}" = 6) transition, while the 852-nm laser was scanned to obtain the two-photon transition spectra with Doppler background; (b) and (d) The 852-nm laser was locked to the 6S$_{1/2}$ (\emph{F} = 4) - 6P$_{3/2}$ (\emph{F}' = 5) transition, while the 698-nm laser was scanned over the Cs 6P$_{3/2}$ (\emph{F}' = 5) - 7D$_{5/2}$  (\emph{F}" = 6, 5, 4) transitions to obtain two-photon transition spectra without Doppler background. }
\label{Fig 5}
\vspace{-0.0in}
\end{figure}

Compared with the spectra obtained by scanning 698-nm coupling laser, the spectrum has a large Doppler background. And when we lock the weak 852 nm probe laser to 6S$_{1/2}$ (\emph{F} = 4) - 6P$_{3/2}$ (\emph{F}' = 5) hyperfine transition, Scanning the 698-nm laser at 6P$_{3/2}$ (\emph{F}' = 5) - 7D$_{5/2}$ transitions to obtain the spectra without Doppler background shown in Fig.5 (b) and (d).

As shown in Fig.5, it can be seen that the resulting linewidth in CP configuration is wider than in the CTP configuration. There exists atomic coherence effect, here EIT, and incoherence effect such as spontaneous emission and DROP. The narrow part of the spectrum due to the coherent process of atoms in the CTP configuration which is caused by the EIT. The wider spectral is contributed by optical double-resonance pumping process. In the CP configuration, DROP, accompanied with spontaneous emission, is a two-photon optical pumping process which is caused by the incoherence effect [33].

\subsection{Effect of the laser beam’s intensity}
\begin{figure}[b]
\vspace{-0.00in}
\centerline{
\includegraphics[width = 86mm]{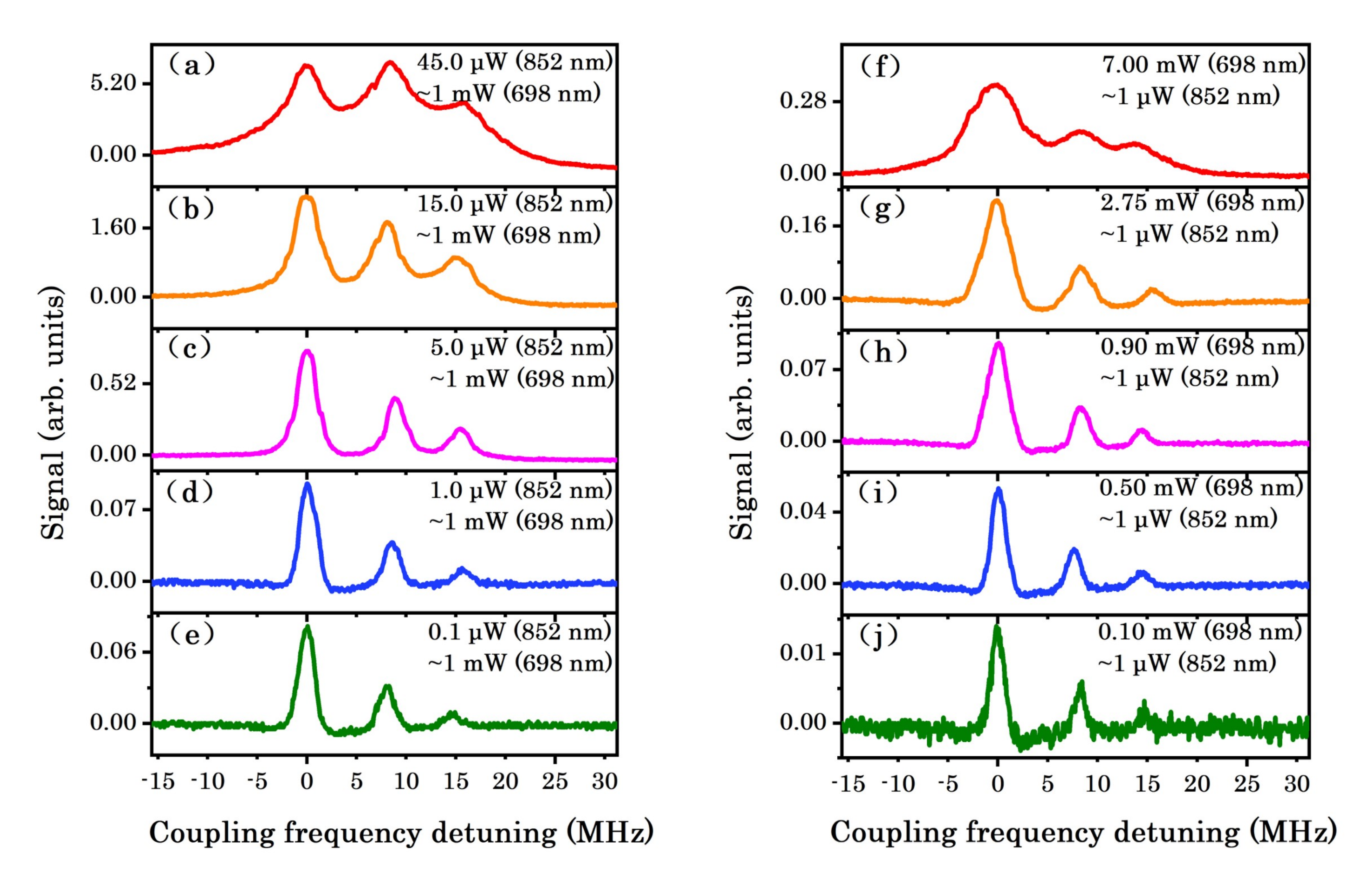}
} \vspace{-0.2in} 
\caption {Evolution of the EIT signals in the CTP configuration with increasing probe laser power (a)-(e) (coupling laser power (f)-(j)). (a)-(e) the 698-nm power is $\sim$1 mW, evolution of the signals with increasing probe laser power; (f)-(j) the 852-nm power is $\sim$1 $\mu$W, evolution of the signals with increasing coupling laser power.}
\label{Fig 6}
\vspace{-0.0in}
\end{figure}
As shown in Fig.6, there are three enhanced transmission signals in the spectrum, and the three peaks from left to right correspond to cesium atom 6P$_{3/2}$ (\emph{F}' = 5) - 7D$_{5/2}$ (\emph{F}" = 6), (\emph{F}" = 5) and (\emph{F}" = 4) hyperfine transitions respectively. The evolution of signals under different probe laser power and coupling laser power is studied. In Fig.6 (a)-(e), when the coupling power is set to 1 mW, with increasing the power of probe beam, the amplitude of EIT-assisted DROP signals increases. When the power of probe beam is set to 1 $\mu$W, with increasing the power of coupling beam, the amplitude of EIT-assisted DROP signals also increases in Fig.6 (f)-(j), the SNR of spectral line decreases obviously when the power of coupling laser power is too weak and is gradually submerged in the noise. With increasing of the coupling or probe beam, the EIT effect becomes stronger. Because the DROP effect mainly depends on the power of 852-nm beam, the optical double resonance effect will widen the spectral linewidth and make it difficult to clearly separate the hyperfine level intervals. In order to clearly distinguish the intervals between three peaks, appropriate powers are selected for measurement in the experiment.

\begin{figure}[h]
\vspace{-0.00in}
\centerline{
\includegraphics[width = 86mm]{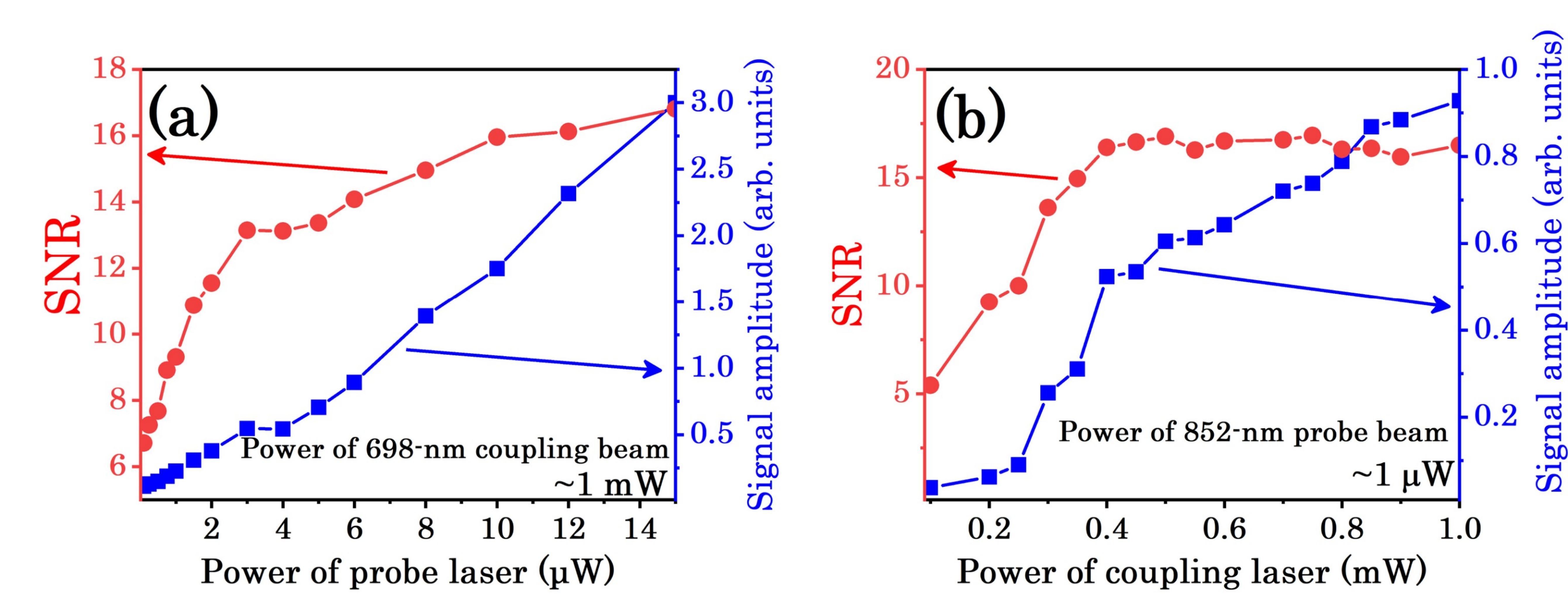}
} \vspace{0.1in} 
\caption {The SNR and amplitude of the cesium 7D$_{5/2}$ (\emph{F}" = 6) spectroscopic signals versus the 852-nm probe beam’s power (a) and the 698-nm coupling beam’s power (b).}
\label{Fig 7}
\vspace{-0.0in}
\end{figure}
We study the variation of signal amplitude and SNR of EIT-assisted DROP spectrum with laser power. It can be seen from Fig.7 The measurement interval we chose is for the purpose of obtaining a spectrum with relatively well SNR and narrow linewidth, so as to make 7D$_{5/2}$ (\emph{F}" = 6, 5, 4) three-peak intervals splitted. To sum up, we choose EIT-assisted DROP spectrum with appropriate power range, all of these advantages make it particularly valuable in measuring the hyperfine splitting intervals and the hyperfine-interaction constants. 

\section{Measurement results and analysis}
In Fig. 8, The frequency calibration is performed by using a electro-optic phase modulator with 30-MHz modulation sideband at 698 nm. The electro-optical phase modulator is driven by the frequency integrator (Agilent Technologies E8257C), and the frequency stability is 1$\times$10$^{-10}$/s. The transmission peak of 7D$_{5/2}$ (\emph{F}" = 6) is used as the reference position for detecting zero detuning of the frequency. 

\begin{figure}[h]
\vspace{-0.00in}
\centerline{
\includegraphics[width = 86mm]{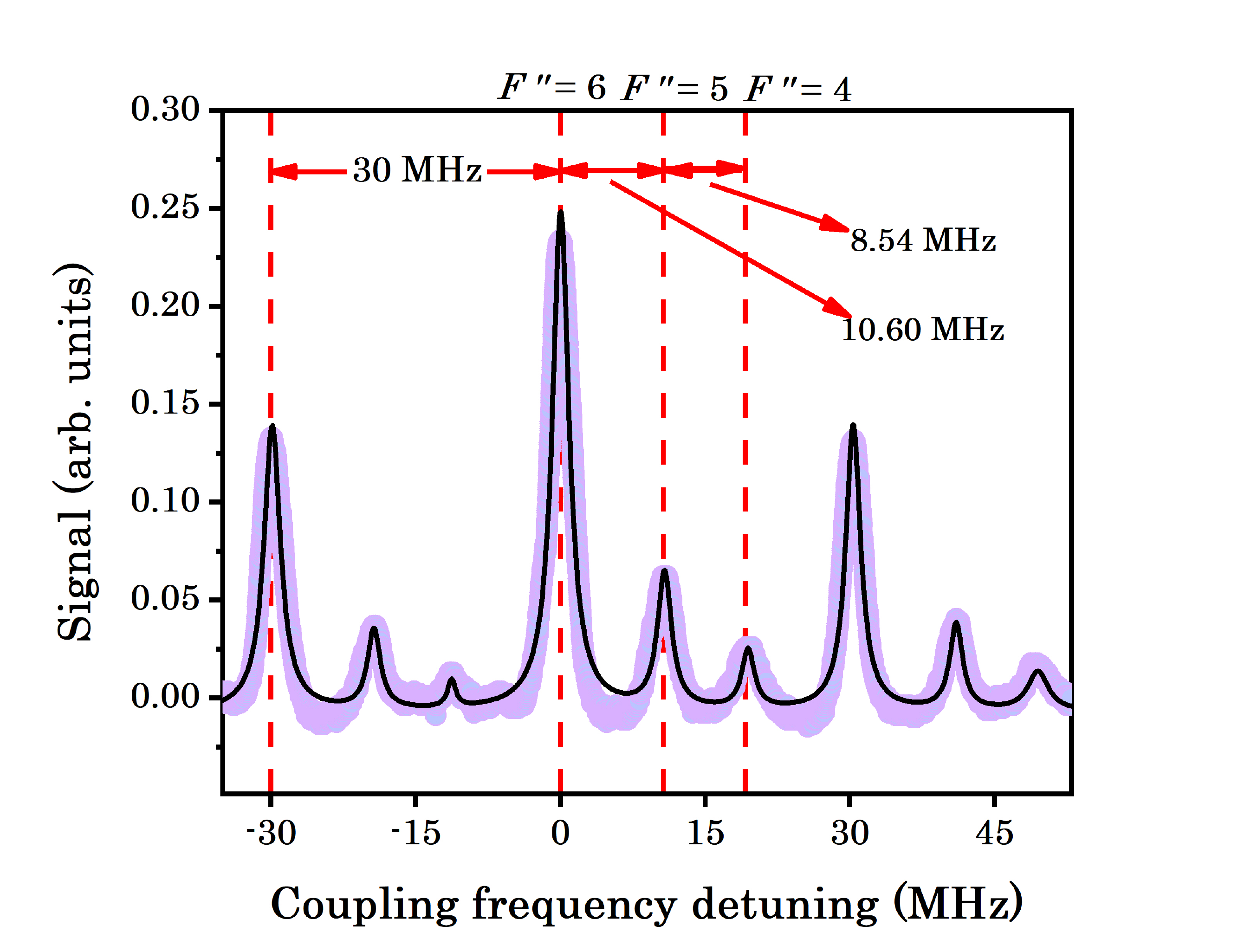}
} \vspace{-0.2in} 
\caption {Calibration of the frequency interval with a 30-MHz radio-frequency phase modulation applied to the coupling laser beam via EOPM. The frequency interval between the carrier (reasonably indicated 0 detuning) and the -1-order sideband (or the +1-order sideband) should be 30 MHz. The solid line displays multi-peak Lorentz fitting, while the light gray curve displays the experimental spectrum.}
\label{Fig 8}
\vspace{-0.0in}
\end{figure}

\begin{figure}[h]
\setlength{\belowcaptionskip}{0.0cm}
\vspace{-0.4in}
\centerline{
\includegraphics[width=90mm]{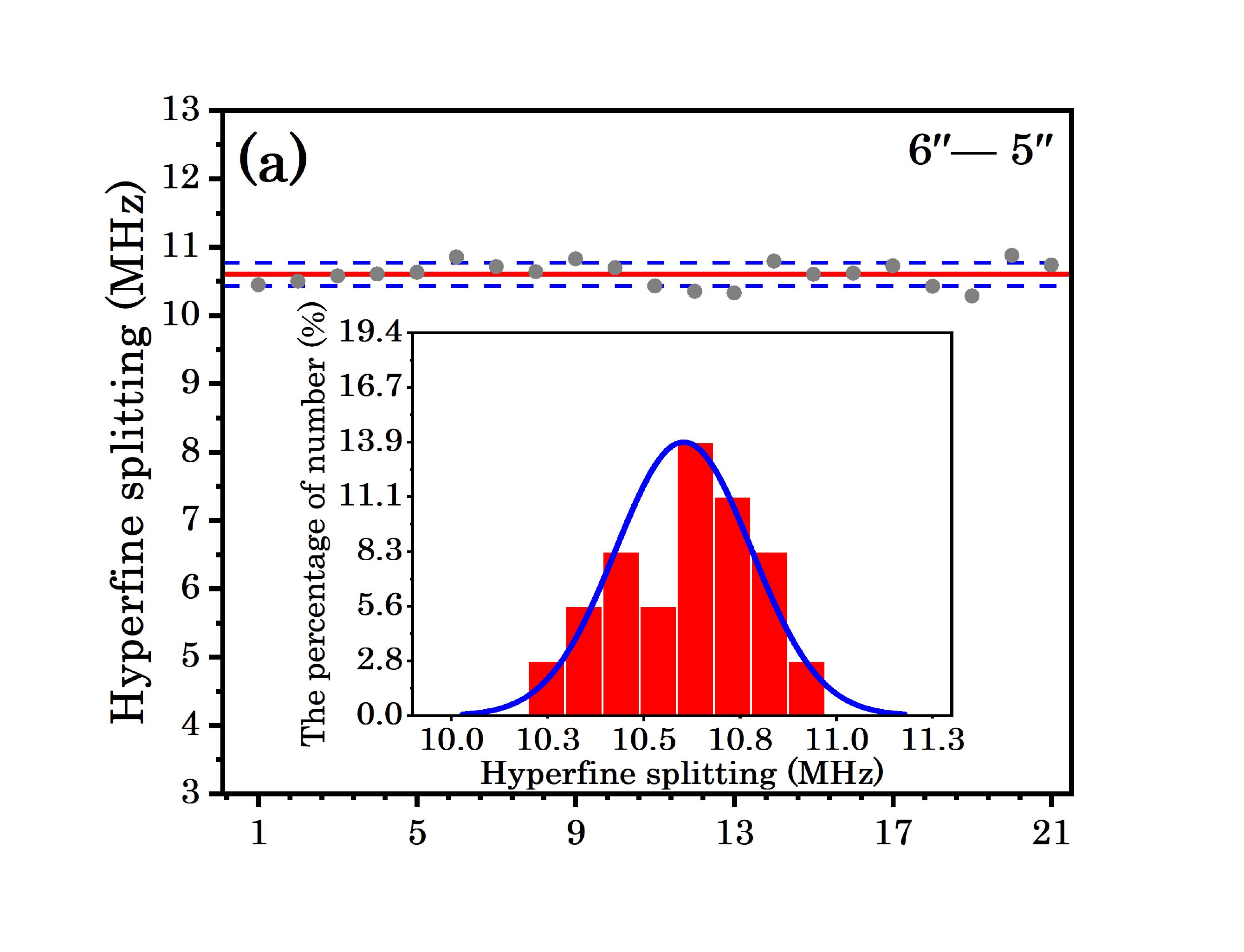}
} \vspace{-0.2in} 
\centerline{
\includegraphics[width=90mm]{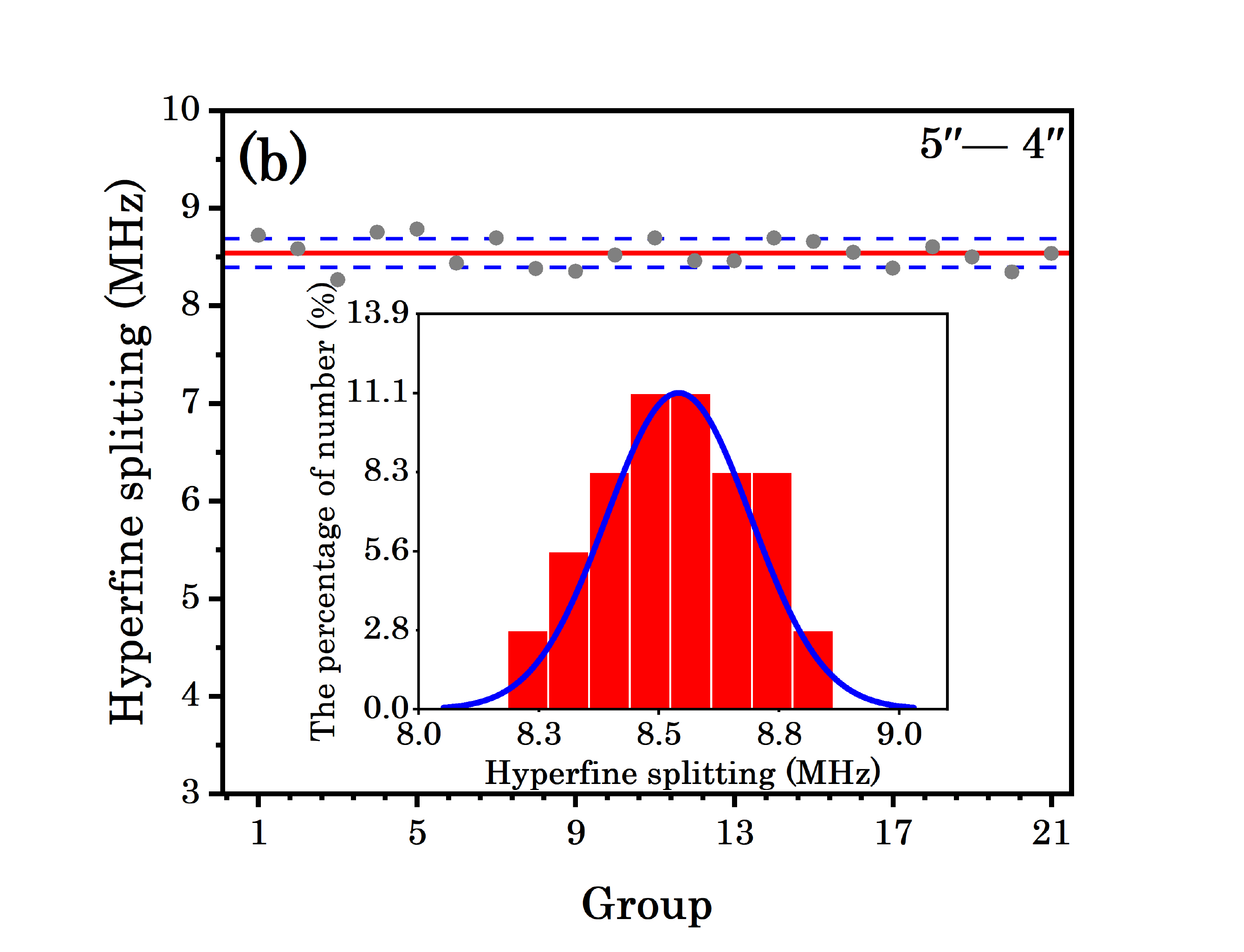}
} \vspace{-0.2in} 
\caption{The measured values of the hyperfine splitting intervals for the Cs 7D$_{5/2}$ excited state. The horizontal red lines stand for the mean values of the hyperfine splitting. The ranges between the two horizontal blue dash lines stand for the statistical errors, and the histograms of the hyperfine splitting intervals are shown as the insets. (a) Hyperfine splitting between (\emph{F}" = 6) and (\emph{F}" = 5) levels of cesium 7D$_{5/2}$ state; (b) Hyperfine splitting between (\emph{F}" = 5) and (\emph{F}" = 4) levels of cesium 7D$_{5/2}$ state.}
\label{Fig 9}
\vspace{0.0in}
\end{figure}

Compared with the methods they used in the reference [27, 30], they used the monochromatic 6S$_{1/2}$ - 7D$_{3/2, 5/2}$ electric quadrupole transition to coupling the ground state and the excited state, which is the amonochromatic two-photon transition. The probability is low and the power of laser is much higher. They also measured the hyperfine splitting intervals of the corresponding states, and their experimental results showed that the spectrum of SNR they obtained were high due to the large splitting intervals of the 7D$_{3/2}$ states. But for the 7D$_{5/2}$ state, the hyperfine splitting intervals are narrow and the resolution is so poor that it is difficult to separate the peaks of 7D$_{5/2}$ (\emph{F}" = 6, 5, 4). As presented in this work, we use the EIT-assisted DROP spectra with well SNR and narrow linewidth, so as to make 7D$_{5/2}$ (\emph{F}" = 6, 5, 4) three-peak intervals splitted. All of these advantages make it valuable in measuring the hyperfine interval and the hyperfine interaction constants. \\

The frequency interval between the carrier frequency and the -1-order sideband serves as the frequency calibration to measure 7D$_{5/2}$ (\emph{F}" = 6, 5, 4) frequency intervals. We use the interval between -1 sideband peak and 7D$_{5/2}$ (\emph{F}" = 6) peak to calibrate the horizontal axis frequency in the Fig.8, so that the frequency intervals between states 7D$_{5/2}$ (\emph{F}" = 6, 5, 4) can be calculated. The frequency of the selected RF signal is 30 MHz, which is close to the interval value of 7D$_{5/2}$ state measured, in order to minimize the influence of the nonlinear effect when scanning laser frequency. Because of the nonlinearity of frequency scanning of grating ceramics when scanning the frequency, the larger the scanning range, the greater the influence of the nonlinear effect. The hyperfine splitting intervals of 7D$_{5/2}$ (\emph{F}" = 6, 5, 4) state are measured and calibrated for many times. The average frequency splitting intervals between cesium 7D$_{5/2}$ (\emph{F}" = 6, 5, 4) hyperfine levels are 10.60(0.17) MHz (5" - 6") and 8.54(0.15) MHz (4" –5"), respectively. Table 1 and Fig.9 show the mean values and errors of the energy level interval obtained through multiple fitting.

The magnetic-dipole hyperfine-interaction constant \textbf{\emph{A}} and the electric-quadrupole hyperfine-interaction constant \textbf{\emph{B}} for the 7D$_{5/2}$ excited state are determined by the splitting interval of the spectrum and can be expressed as follows:
\begin{equation}
{{HFS}_{6"-5"}} = {6}{\textbf{\emph{A}}}+\frac{18}{35}{\textbf{\emph{B}}} = -10.60(0.17) MHz,
\end{equation}
\begin{equation}
{{HFS}_{5"-4"}} = {5}{\textbf{\emph{A}}}+\frac{3}{70}{\textbf{\emph{B}}} = -8.54(0.15) MHz.
\end{equation}

According to the equations (4) and (5), the magnetic-dipole hyperfine-interaction constant \textbf{\emph{A}} = -1.70(0.03) MHz and the electric-quadrupole hyperfine-interaction constant \textbf{\emph{B}} = -0.77(0.58) MHz of the 7D$_{5/2}$ excited state can be calculated. The results are in good agreement with the previous reported results [27-29, 34], as shown in Table 2. The experimental error mainly comes from the spectral resolution limitation caused by the wide linewidth of the obtained spectrum and the accuracy of the function generator, which driving the electro-optical phase modulator. 
\linespread{1.3}
\begin{table}[h]

\caption{\label{tab:table1}%
Measured values of hyperfine splitting of Cs 7D$_{5/2}$ state}
\begin{ruledtabular}

\begin{tabular}{lcdr}
\textrm{Hyperfine Splitting}&
\textrm{Measured Splitting Values (MHz)}\\
\colrule
7D$_{5/2}$ (\emph{F}" = 6) - (\emph{F}" = 5)\\HFS$_{ 6"-5"}$ & -10.60(0.17)\\
7D$_{5/2}$ (\emph{F}" = 5) - (\emph{F}" = 4)\\HFS$_{ 5"-4"}$ & -8.54(0.15)\\
\end{tabular}
\end{ruledtabular}
\end{table}

\linespread{1.3}
\begin{table}[h]
\caption{\label{tab:table2}%
Magnetic-dipole (\textbf{\emph{A}}) and electric-quadrupole (\textbf{\emph{B}}) hyperfine-interaction constants of Cs 7D$_{5/2}$ state.}
\begin{ruledtabular}
\begin{tabular}{cccccccccccc}
\textrm{\textbf{\emph{A}} (MHz)}&\textrm{\textbf{\emph{B}} (MHz)}& 
\textrm{Experiment/Ref/Year}\\
\colrule
-1.70(0.03) & -0.77(0.58) &  \multicolumn{1}{p{3.5cm}}{This work}\\
-1.79(0.05) & +1.05(0.29) &  \multicolumn{1}{p{3.5cm}}{Ref. [29] (2020)}\\
-1.8 (0.05) & +1.01(1.06) &  \multicolumn{1}{p{3.5cm}}{Ref. [27] (2011)}\\
-1.717(0.015) & -0.18(0.52) &  \multicolumn{1}{p{3.5cm}}{Ref. [28] (2010)}\\
-1.56(0.09) & - &  \multicolumn{1}{p{3.5cm}}{Ref. [34] (2007)}\\
\end{tabular}
\end{ruledtabular}
\end{table}

\section{Conclusion}
In conclusion, we measured the hyperfine splitting intervals and derived the hyperfine-interaction constants of cesium 7D$_{5/2}$ excited state. Cs atoms were excited from the ground state 6S$_{1/2}$ to the excited state 7D$_{5/2}$ (\emph{F}" = 6, 5, 4)through the ladder-type 6S$_{1/2}$ - 6P$_{3/2}$ - 7D$_{5/2}$ three-level system. The OODR and EIT-assisted DROP spectroscopic schemes were compared, and the EIT was employed to get much better SNR and higher resolution. Also the effect of 852-nm probe beam’s linewidth and optical intensity of 852-nm probe beam and 698-nm coupling beam upon the spectroscopic amplitude were investigated. After selection of spectroscopic scheme and optimization of experimental parameters, the Cs 7D$_{5/2}$ state’s hyperfine splitting intervals between (\emph{F}" = 6), (\emph{F}" = 5), and (\emph{F}" = 4) hyperfine folds have been measured (HFS$_{ 6"-5"}$ = -10.60(0.17) MHz and HFS$_{5"-4"}$ = -8.54(0.15) MHz) by introducing EOPM to provide modulation sidebands as frequency calibration. Furthermore, the hyperfine-interaction constants (\textbf{\emph{A}} = -1.70(0.03) MHz is the magnetic-dipole constant and \textbf{\emph{B}} = -0.77(0.58) MHz is the electric-quadrupole constant) have been derived for the Cs 7D$_{5/2}$ state. The measurement results are compared and consistent with the previous measurement results, and these measurement results will provide more data for references to the further relevant theoretical and experimental works.

Considering that the frequency locking precision of 852-nm laser needs to be improved, the nonlinearity of 698-nm coupling frequency scanning, and the fluctuation of atomic number density caused by the temperature fluctuation of cesium atom vapor due to the absence of temperature control, the accuracy of hyperfine splitting intervals needs to be further improved. The contributions made here have wide applications. The combination of atomic coherence and optical pumping spectroscopy allows that the  EIT-assisted DROP spectroscopic scheme has much better SNR and higher frequency resolution, which is of great significance for precise measurement of spectral fields. On the other hand, it provides an important reference value for measuring the internal structure of atoms to deal with the problem of PNC and other related works. 

\begin{acknowledgments}
This work is partially supported by the National Key R\&D Program of China (Grant No. 2017YFA0304502), the National Natural Science Foundation of China (Grant Nos. 11774210, 11974226, 61875111, and 61905133), and Shanxi Provincial 1331 Project for Key Subjects Construction. 
\end{acknowledgments}
\linespread{1}


\begin{thebibliography}{99}
\bibitem{ref-journal1}R. Li, Y. Wu, Y. Rui, B. Li, Y. Jiang, L. Ma, and H. Wu, Absolute frequency measurement of $^{6}$Li D lines with khz-Level uncertainty, {\em Phys. Rev. Lett.} {\bf 2020}, {\em 124}, 063002.
\bibitem{ref-journal2}T. Ray, R. K. Gupta1, V. Gokhroo1, J. L. Everett, T. Nieddu, K. S Rajasree and S. N. Chormaic, Observation of the  $^{87}$Rb 5S$_{1/2}$ to 4D$_{3/2}$ electric quadrupole transition at 516.6 nm mediated via an optical nanofibre. {\em New J. Phys.} {\bf 2020}, {\em 22}, 062001.
\bibitem{ref-journal3}J. Yuan, C. Wu, Y. Li, L. Wang, Y. Zhang, L. Xiao, S. Jia, Controllable electromagnetically induced grating in a cascade-type atomic system, {\em Front. Phys.} {\bf 2019}, {\em 14}, 52603.
\bibitem{ref-journal4}X. Zheng, Y. Sun, J. Chen, W. Jiang, K. Pachucki, and S. Hu, Measurement of the frequency of the 2$^{3}$S - 2$^{3}$P transition of $^{4}$He, {\em Phys. Rev. Lett.} {\bf 2017}, {\em 119}, 263002.
\bibitem{ref-journal5}V. Gerginov, A. Derevianko, and C. E. Tanner, Bservation of the nuclear magnetic octupole moment of Cs 133. {\em Phys. Rev. Lett.} {\bf 2003}, {\em 91}, 072501.
\bibitem{ref-journal6}C. S. Wood, S. C. Bennett, D. Cho, B. P. Masterson, J. L. Roberts, C. E. Tanner, and C. E. Wieman, Measurement of parity nonconservation and an anapole moment in cesium. {\em Science} {\bf 1997}, {\em 275}, 1759.
\bibitem{ref-journal7}S. G. Porsev, K. Beloy, and A. Derevianko, Precision determination of electroweak coupling from atomic parity violation and implications for particle physics. {\em Phys. Rev. Lett. } {\bf 2009}, {\em 102}, 181601.
\bibitem{ref-journal8}M. Y. Kuchiev and V. V. Flambaum, Influence of perturbations on the electron wave function inside the nucleus. {\em J. Phys. B: At. Mol. Opt. Phys.} {\bf 2002}, {\em 35}, 4101.
\bibitem{ref-journal9}M. G. Kozlov, S. G. Porsev, and I. I. Tupitsyn, High accuracy calculation of 6S-7S parity nonconserving amplitude in Cs. {\em Phys. Rev. Lett.} {\bf 2001}, {\em 86}, 3260.
\bibitem{ref-journal10}B. K. Sahoo, G. Gopakumar, R. K. Chaudhuri, B. P. Das, H. Merlitz, U. S. Mahapatra, and D. Mukherjee, Magnetic-dipole hyperfine interactions in $^{137}$Ba$^{+}$ and the accuracies of the neutral weak interaction matrix elements.{\em Phys. Rev. A} {\bf 2003}, {\em 68}, 040501.
\bibitem{ref-journal11}B. K. Sahoo and B. P. Das, Constraints on new physics from an improved calculation of parity violation in $^{133}$Cs. {\em arXiv:2008.08941v1} {\bf 2020}.
\bibitem{ref-journal12}M. S. Safronova and C. W. Clark, Inconsistencies between lifetime and polarizability measurements in Cs. {\em Phys. Rev. A} {\bf 2004}, {\em 69}, 361.
\bibitem{ref-journal13}M. S. Safronova, U. I. Safronova, and C. W. Clark, Magicwavelengths, matrix elements, polarizabilities, and lifetimes of Cs, {\em Phys. Rev. A} {\bf 2016}, {\em 94}, 012505.
\bibitem{ref-journal14}V. A. Dzuba, V. V. Flambaum, and J. S. Ginges, Calculations of parity-nonconserving s-d amplitudes in Cs, Fr, Ba$^{+}$, and Ra$^{+}$. {\em Phys. Rev. A} {\bf 2001}, {\em 63}, 21.
\bibitem{ref-journal15}K. Heshamia, D. G. Englanda, P. C. Humphreysb, P. J. Bustarda, V. M. Acostac, J. Nunnb, and B. J. Sussmana, Quantum memories: emerging applications and recent
advances. {\em J Mod Opt.} {\bf 2016}, {\em 63}, 2005.
\bibitem{ref-journal16}B. D. Yang, Q. B. Liang, J. He, T. C. Zhang, and J. M. Wang, Narrow-linewidth double-resonance optical pumping spectrum due to electromagnetically induced transparency in ladder-type inhomogeneously broadened media. {\em Phys. Rev. A} {\bf 2010}, {\em 81}, 043803.
\bibitem{ref-journal17}A. G. Sinclair, B. D. Mcdonald, E. Riis, and G. Duxbury, Double resonance spectroscopy of laser-cooled Rb atoms. {\em Opt. Commun.} {\bf 1994}, {\em 106}, 207.
\bibitem{ref-journal18}P. Fendel, S. D. Bergeson, Th. Udem, and T. W. Hänsch, Two-photon frequency comb spectroscopy of the 6S-8S transition in cesium. {\em Opt. Lett.} {\bf 2007}, {\em 32}, 701.
\bibitem{ref-journal19}J. Yuan, S. Dong, C. Wu, L. Wang, L. Xiao, and S. Jia, Optically tunable grating in a {\em V} + {$\Xi$} conguration involving a Rydberg state. {\em Opt. Express} {\bf 2020}, {\em 28}, 23820.
\bibitem{ref-journal20}T. T. Grove, V. Sanchez-Villicana, B. C. Duncan, S. Maleki, and P. L. Gould, Two-photon two-color diode laser spectroscopy of the Rb 5D$_{5/2}$ state. {\em Physica Scripta} {\bf 1995}, {\em 52}, 271.
\bibitem{ref-journal21}M. Auzinsh, K. Blushs, R. Ferber, F. Gahbauer, A. Jarmola, and M. Tamanis, Electric field induced hyperfine level-crossings in (nD) Cs at two-step laser excitation: experiment and theory. {\em Opt. Commun.} {\bf 2006}, {\em 264}, 333.
\bibitem{ref-journal22}J. Wang, H. F. Liu, G. Yang, B. D. Yang, and J. M. Wang, Determination of the hyperfine structure constants of the $^{87}$Rb and $^{85}$Rb 4D$_{5/2}$ state and the isotope hyperfine anomaly. {\em Phys. Rev. A} {\bf 2014}, {\em 90}, 052505.
\bibitem{ref-journal23}J. Wang, H. F. Liu, B. D. Yang, J. He, and J. M. Wang, Determining the hyperfine structure constants of cesium 8S$_{1/2}$ state aided by atomic coherence. {\em Meas. Sci. Technol.} {\bf 2014}, {\em 25}, 035501.
\bibitem{ref-journal24}G. Yang, J. Wang, and J. M. Wang, Determination of the hyperfine coupling constants of the 5D$_{5/2}$ state of $^{85}$Rb atoms by using high signal-to-noise ratio electromagnetically-induced transparency spectra.{\em Acta Physica Sinica} {\bf 2017}, {\em 66}, 103201.(in chinese)
\bibitem{ref-journal25}Y. H. He, J. B. Fan, L. P. Hao, Y. C. Jiao, and J. M. Zhao, Precise measurement of hyperfine structure of cesium 7S$_{1/2}$ excited state. {\em Appl. Sci.} {\bf 2020}, {\em 10}, 525.
\bibitem{ref-journal26}B. R. Bulos, R. Gupta, G. Moe, and P. Tsekeris, Hyperfine structure determination of the 7D$_{5/2}$ state of $^{133}$Cs. {\em Phys. Rev. A} {\bf 1976}, {\em 55}, 407.
\bibitem{ref-journal27}Y. C. Lee, Y. H. Chang, Y. Y. Chang, Y. Y. Chen, C. C. Tsai, and H. C. Chui, Hyperfine coupling constants of cesium 7D states using two-photon spectroscopy. {\em  Appl. Phys. B} {\bf 2011}, {\em 105}, 391.
\bibitem{ref-journal28}J. E. Stalnaker, V. Mbele, V. Gerginov, T. M. Fortier, and C. E. Tanner, Femto-second frequency comb measurement of absolute frequencies and hyperfine coupling constants in cesium vapor. {\em  Phys. Rev. A} {\bf 2010}, {\em 81}, 043840.
\bibitem{ref-journal29}S. D. Wang, J. P. Yuan, L. R. Wang, L. T. Xiao, and S. T. Jia, Investigation on the Cs 6S$_{1/2}$ to 7D electric quadrupole transition via monochromatic two-photon process at 767 nm {\em arXiv:2008.09739v1} {\bf 2020}.
\bibitem{ref-journal30}A. Ramos, R. Cardman, and G. Raithel, Measurement of the hyperfine coupling constant for nS$_{1/2}$ Rydberg states of $^{85}$Rb. {\em Phys. Rev. A} {\bf 2019}, {\em 100}, 062515.
\bibitem{ref-journal31}G. Yang, J. Wang, B. D. Yang, and J. M. Wang, Determination of the hyperfine coupling constant of the cesium 7S$_{1/2}$ state. {\em Laser Phys. Lett.} {\bf 2016}, {\em 13}, 085702.
\bibitem{ref-journal32}B. D. Yang, J. Gao, Q. B. Liang, J. Wang, T. C. Zhang, and J. M. Wang, Double-resonance optical-pumping effect and ladder-type electromagnetically induced transparency signal without doppler background in cesium atomic vapour cell. {\em  Chinese Phys. B} {\bf 2011}, {\em 20}, 044202.
\bibitem{ref-journal33}J. Gea-Banacloche, Y. Q. Li, S. Z. Jin, and M. Xiao, Electromagnetically induced transparency in ladder-type inhomogeneously broadened media: theory and experiment. {\em  Phys. Rev. A} {\bf 1995}, {\em 51}, 576.
\bibitem{ref-journal34}M. Auzinsh, K. Blushs, R. Ferber, F. Gahbauer, A. Jarmola, and M. S. Safronova, Level-crossing spectroscopy of the 7, 9, and 10D$_{5/2}$ states of $^{133}$Cs and validation of relativistic many-body calculations of the polarizabilities and hyperfine constants. {\em  Phys. Rev. A} {\bf 2007}, {\em 75}, 22502.

\end{thebibliography}
\end{document}